\documentclass[11pt,superscriptaddress,showpacs,aps,preprint]{revtex4}

\usepackage[latin9]{inputenc}

\usepackage{amsmath}
\usepackage{amssymb}
\usepackage{graphicx}
\usepackage{color}

\makeatletter

\usepackage{slashed}

\def\as{a\!\!\!/}
\def\ks{k\!\!\!/}
\def\ls{l\!\!\!/}
\def\ps{p\!\!\!/}

\def\bs{b\!\!\!/}

\newcommand{\be}{\begin{equation}}
\newcommand{\ee}{\end{equation}}
\newcommand{\en}{\end{equation}}
\newcommand{\ba}{\begin{eqnarray}}
\newcommand{\ea}{\end{eqnarray}}
\newcommand{\bea}{\begin{eqnarray}}
\newcommand{\eea}{\end{eqnarray}}
\newcommand{\pa}{\partial}
\newcommand{\bq}{\begin{eqnarray}}
\newcommand{\eq}{\end{eqnarray}}
\newcommand{\kbruto}{\hbox{$k \!\!\!{\slash}$}}
\newcommand{\pbruto}{\hbox{$p \!\!\!{\slash}$}}

\makeatother

\begin{document}

\title {Higher-order one-loop contributions in Lorentz-breaking QED}

\author{A. P. Baeta Scarpelli}
\email{scarpelli.abps@dpf.gov.br}
\affiliation{Centro Federal de Educa\c{c}\~ao Tecnol\'ogica - MG \\
Avenida Amazonas, 7675 - 30510-000 - Nova Gameleira - Belo Horizonte
-MG - Brazil}

\author{L. C. T. Brito}
\email{lcbrito@dfi.ufla.br}
\affiliation{Departamento de Física, Universidade Federal de Lavras, Caixa Postal 3037,
37200-000, Lavras, MG, Brasil}

\author{J. C. C. Felipe}
\email{jean.cfelipe@ufvjm.edu.br}
\affiliation{Instituto de Engenharia, Ciência e Tecnologia, Universidade Federal dos Vales do Jequitinhonha e Mucuri, Avenida Manoel Bandeira, 460 - 39440-000 - Veredas - Janaúba - MG- Brazil}

\author{J. R. Nascimento}
\email{jroberto@fisica.ufpb.br}
\affiliation{Departamento de Física, Universidade Federal da Paraíba, Caixa Postal 5008,
58051-970 João Pessoa, Paraíba, Brazil}

\author{A. Yu. Petrov}
\email{petrov@fisica.ufpb.br}
\affiliation{Departamento de Física, Universidade Federal da Paraíba, Caixa Postal 5008,
58051-970 João Pessoa, Paraíba, Brazil}

\pacs{11.30.Cp}

\begin{abstract}
We calculate higher-order quantum contributions in different Lorentz-violating parameters to the gauge sector of the extended QED. As a result of this one-loop calculation, some terms which do not produce first-order corrections, contribute with nontrivial gauge-invariant second-order quantum inductions.

\end{abstract}

\maketitle

\section{Introduction}
\label{I}

An important issue when the impacts of a possible violation of Lorentz symmetry are studied consists in the investigation of Lorentz-breaking extensions of known field-theory models \cite{kostelecky1, kostelecky2}. A great list of various extensions of Quantum Electrodynamics (QED), scalar field theory and gravity is presented in \cite{Kostel}. A very important subject of study is the quantum dynamics of such theories, since the inclusion of these parts in the classical action may cause the radiative induction of new terms. The most well-known result is the perturbative generation of the Carroll-Field-Jackiw (CFJ) term \cite{CFJ} when an axial term is included in the fermionic sector of the extended QED, which was discussed for the first time in \cite{JK}. Further, many aspects of calculations of the quantum induction of the CFJ term have been discussed in dozens of papers. The articles treated issues like the ambiguity of the induced term, its finite-temperature aspects, its non-Abelian generalization, the proper-time approach of the calculation and many implications (see f.e. \cite{CS4}-\cite{CS13} and references therein). Further, the nonminimal interaction has been used to generate the CPT-even aether-like term \cite{aether} as well as the CFJ one \cite{aether2}.

At the same time, it is well-known that the number of possible Lorentz-breaking extensions, even after imposing the restrictions of renormalizability and absence of higher derivatives, is very large \cite{Kostel}. It is to be noted, however, that only a small part of these corrections was considered at the perturbative level (see also \cite{MarizHD}). Besides, in most cases the investigation was restricted to the first order in Lorentz-breaking parameters. While it is reasonable (remind that the Lorentz-breaking parameters are very small \cite{datatables}), the problem of possible higher-order Lorentz-breaking corrections certainly deserves attention. This point of view is reinforced by the observation that the first-order correction is null for some of these Lorentz-breaking tensors, see f.e. \cite{Kost2001a}. It is relevant to investigate if this behavior is preserved to all orders due to a deeper reason or if this null value is only eventual.

In this paper, we consider the second-order corrections in the parameters $e^{\mu}$, $f^{\mu}$, $a_{\mu}$, $g_{\mu\nu\lambda}$ and $H_{\mu\nu}$  of the extended Quantum Electrodynamics, which were not calculated up to now \cite{Kostel}. The paper is organized as follows: in section II, we write down a generic Lorentz-breaking extension of QED and review the first-order quantum corrections. In section III, we calculate the second-order quantum corrections in the Lorentz-breaking tensors. The section IV is left for the summary, in which the results are discussed.

\section{A generic Lorentz-breaking extension of QED and first-order quantum corrections}
\label{II}
The most generic Lorentz-breaking renormalizable extension of QED containing no higher derivatives is given by the following Lagrangian \cite{Kost2001a}:
\bea
\label{genrenmod}
{\cal L}=\bar{\psi}(i\Gamma^{\nu}D_{\nu}-M)\psi-\frac{1}{4}F_{\mu\nu}F^{\mu\nu}-\frac{1}{4}\kappa_{\mu\nu\lambda\rho}F^{\mu\nu}F^{\lambda\rho}+\frac{1}{2}\epsilon_{\mu\nu\lambda\rho}k^{\mu}A^{\nu}F^{\lambda\rho},
\eea
in which
\be
\Gamma^{\nu}=\gamma^{\nu}+c^{\mu\nu}\gamma_{\mu}+d^{\mu\nu}\gamma_{\mu}\gamma_5+e^{\nu}+if^{\nu}\gamma_5+\frac{1}{2}g^{\lambda\mu\nu}\sigma_{\lambda\mu}
\ee
and
\be
M=m+a_{\mu}\gamma^{\mu}+b_{\mu}\gamma^{\mu}\gamma_5+\frac{1}{2}H^{\mu\nu}\sigma_{\mu\nu}.
\ee
Besides, $D_{\mu}=\pa_{\mu}-iqA_{\mu}$ is the simplest covariant derivative, being $q$ the coupling constant, and $\kappa_{\mu\nu\lambda\rho}$, $k^{\mu}$, $a^{\mu}$, $b^{\mu}$, $c^{\mu\nu}$, $d^{\mu\nu}$, $e^{\mu}$, $f^{\mu}$, $g^{\lambda\mu\nu}$ and $H^{\mu\nu}$ are constant (pseudo)tensors, which are responsible for the Lorentz-symmetry violation.

We are interested in the study of the quantum corrections to the gauge sector of this theory. To consider quantum corrections to the gauge sector, it is sufficient to treat the photon as an external field, and to integrate out the spinorial field. The corresponding fermionic determinant is evaluated up to the necessary order in the couplings, and, in the Lorentz-breaking context, to the necessary order in Lorentz-breaking parameters.  For our purpose, the most interesting constant vector and tensor parameters are those ones contributing to $\Gamma^{\nu}$ and $M$.

Some of these vectors or tensors have been intensively studied, such as $b_{\mu}$, used to generate the CFJ term (see f.e. \cite{CFJ,JK}). Besides, $c_{\mu\nu}$ and $d_{\mu\nu}$ were discussed in the context of the extension of the ABJ-anomaly in \cite{Arias} (their all-order one-loop contributions to this anomaly have been obtained), and, in \cite{Maluf}, the aether-like contributions up to the third order in $c_{\mu\nu}$ were calculated. Considering the first-order correction in $d_{\mu\nu}$, it can be non-null only if one has $d_{\mu\nu}=C\eta_{\mu\nu}$. Indeed, in the induction of the aether-like term, the obtained tensor $\kappa_{\mu\nu\lambda\rho}$ has only one possible structure in first order in $d_{\mu\nu}$, given by $\kappa_{\mu\nu\lambda\rho}=d_{\mu}^{\alpha}\epsilon_{\alpha\nu\lambda\rho}$ (the Levi-Civita symbol emerges due the fact that $d_{\mu\nu}$ is accompanied by a $\gamma_5$ matrix). However, this form of the $\kappa_{\mu\nu\lambda\rho}$ possesses the necessary symmetry only if $d_{\mu}^{\nu}\propto\delta_{\mu}^{\nu}$, and this case yields a trivial result. Concerning the tensor $g_{\mu\nu\lambda}$ its lower (first-order) non-zero impact in the gauge sector was calculated in \cite{MarizHD}.

Therefore, we are left with problem of the evaluation of quantum corrections involving the remaining parameters, $e^{\mu}$, $f^{\mu}$, $a_{\mu}$ and $H_{\mu\nu}$. It is not difficult to show that their first-order contributions vanish (some preliminary discussions on this fact are presented in \cite{Scarpanom}). Indeed, $e_{\mu}$ and $a_{\mu}$ are vectors and, hence, cannot substitute the axial vector $b_{\mu}$ in the first possible gauge invariant contribution, that is, in the CFJ term, $\epsilon^{\mu\nu\lambda\rho}b_{\mu}A_{\nu}F_{\lambda\rho}$. Considering the first-order contribution proportional to $H_{\mu\nu}$, it would have the form $H_{\mu\nu}F^{\mu\alpha}F^{\nu}_{\phantom{\nu}\alpha}$, which identically vanishes.

On the other hand, the first-order contribution proportional to $f^{\mu}$ yields
\bea
\Gamma_f=\frac{iq^2}{2}A_{\mu}(-p)\Pi^{\mu\nu}_f(p)A_{\nu}(p),
\eea
where
\bea
\Pi^{\mu\nu}_f(p)&=&{\rm tr}\int\frac{d^4k}{(2\pi)^4}\Big[\gamma^{\mu}\frac{1}{\ks-m}\gamma_5(f\cdot k)\frac{1}{\ks-m}\gamma^{\nu}\frac{1}{\ks+\ps-m}+\nonumber\\
&+&\gamma^{\mu}\frac{1}{\ks-m}\gamma^{\nu}\frac{1}{\ks+\ps-m}\gamma_5(f\cdot (k+p))\frac{1}{\ks+\ps-m}
\Big].
\eea
The contributions come from the two possible bubble graphs with one insertion of $f^\mu$. For performing the evaluation of the two terms, one will have to deal, respectively, with the traces,
\be
{\rm tr}\{\gamma^\mu(\kbruto+m)\gamma_5(\kbruto+m)\gamma^\nu(\kbruto+\pbruto+m)\}
\ee
and
\be
{\rm tr}\{\gamma^\mu(\kbruto+m)\gamma^\nu(\kbruto+\pbruto+m)\gamma_5(\kbruto+\pbruto+m)\},
\ee
which are null, as one can easily verify.

Concerning the contributions in which $e^{\mu}$, $f^{\mu}$ and $a_{\mu}$ are contracted not to the derivative but to the $A_{\mu}$ field, we note that the results of first-order in these parameters, proportional to $(e\cdot A)$ (or $f\cdot A$ or $a\cdot A$), are forbidden by gauge invariance. All these arguments match the conclusions of \cite{Kostel}.

\section{Second-order quantum corrections}

Now, we go to the key point of our paper. Being the first order contribution in some parameters null, a natural question is whether these results will be maintained at higher orders or not. Besides, if not, it is relevant to obtain the lowest order non-null contributions in these parameters. Hence, we now perform the calculation of the one-loop second-order corrections in these Lorentz-breaking tensors.

We first discuss the second-order correction in $a_{\mu}$. We have three ways to perform two insertions of $a^\mu$ in the bubble diagram: two possibilities of two insertions in the same internal line; and one possibility of one insertion in each internal line. Then, we have
\begin{eqnarray}
S_{AA}^{(a)}(p)&=&\frac{ie^2}{2}\int\frac{d^4l}{(2\pi)^4}{\rm tr}\Big(\gamma^{\mu}\frac{1}{\ls-m}\gamma^{\nu}\frac{1}{\ls+\ps-m}
\as\frac{1}{\ls+\ps-m}\as\frac{1}{\ls+\ps-m}+\nonumber\\
&+&
\gamma^{\mu}\frac{1}{\ls-m}\as\frac{1}{\ls-m}
\gamma^{\nu}\frac{1}{\ls+\ps-m}\as\frac{1}{\ls+\ps-m}+\nonumber\\
&+&
\gamma^{\mu}\frac{1}{\ls-m}\as\frac{1}{\ls-m}
\as\frac{1}{\ls-m}\gamma^{\nu}\frac{1}{\ls+\ps-m}\Big)
A_{\mu}(-p)A_{\nu}(p).
\label{eea}
\end{eqnarray}
This expression is similar to that one studied in \cite{aether2}, but involves $\as$ instead of $\bs\gamma_5$. It must be expanded up to the second order in the external momentum $p$. Straightforward calculations show that the result for this contribution is zero. From the formal viewpoint, it is related with the fact that if we consider the action (\ref{genrenmod}), the gauge transformation $A_{\mu}\to A_{\mu}+a_{\mu}$ allows to rule out the vector $a_{\mu}$ already at the classical level. Besides, this transformation has no implications in the quantum computations, since the corresponding Jacobian is 1. Hence, we can go beyond and affirm that the corrections in all orders in $a^\mu$ are null.

Let us consider now the second-order contribution in $H_{\mu\nu}$. Again, we have three possibilities of two insertions of $H_{\mu \nu}$ in the internal lines of a bubble diagram. Then, we have
\begin{eqnarray}
S_{AA}^{(H)}(p)&=&\frac{ie^2}{2}\int\frac{d^4l}{(2\pi)^4}{\rm tr}\Big(\gamma^{\mu}\frac{1}{\ls-m}\gamma^{\nu}\frac{1}{\ls+\ps-m}
(\frac{1}{2}H^{\alpha\beta}\sigma_{\alpha\beta})\frac{1}{\ls+\ps-m}(\frac{1}{2}H^{\gamma\delta}\sigma_{\gamma\delta})\frac{1}{\ls+\ps-m}+\nonumber\\
&+&
\gamma^{\mu}\frac{1}{\ls-m}(\frac{1}{2}H^{\alpha\beta}\sigma_{\alpha\beta})\frac{1}{\ls-m}
\gamma^{\nu}\frac{1}{\ls+\ps-m}(\frac{1}{2}H^{\gamma\delta}\sigma_{\gamma\delta})\frac{1}{\ls+\ps-m}+\nonumber\\
&+&
\gamma^{\mu}\frac{1}{\ls-m}(\frac{1}{2}H^{\alpha\beta}\sigma_{\alpha\beta})\frac{1}{\ls-m}
(\frac{1}{2}H^{\gamma\delta}\sigma_{\gamma\delta})\frac{1}{\ls-m}\gamma^{\nu}\frac{1}{\ls+\ps-m}\Big)
A_{\mu}(-p)A_{\nu}(p).
\label{eeh}
\end{eqnarray}
Expanding this expression up to the second order in the external momentum $p$ and performing straightforward calculations, one finds that this contribution also yields zero result. This fact can be explained by the following argument.  First, the only gauge invariant contributions of second order both in $H_{\mu\nu}$ and in derivatives have the form $(H^{\mu\nu}F_{\mu\nu})^2$ and $H^{\mu\nu}H^{\rho\sigma}F_{\mu\rho}F_{\nu\sigma}$. However, it is simple to see from the calculations that these terms do not arise.
Besides, all other contributions of second order in $H_{\mu\nu}$, to be consistent with the gauge invariance, should have the structure $A_{\mu}P^{\mu\nu}H_{\nu\alpha}H^{\alpha\beta}P_{\beta\nu}A^{\nu}$, where $P^{\mu\nu}=\eta^{\mu\nu}\Box-\partial^{\mu}\partial^{\nu}$ is a transverse projector. However, this term is already of fourth order in derivatives and, since our spinors are massive, this order will not be reduced by factors of $\Box^{-1}$. Hence, gauge invariant terms of second-order both in derivatives and in $H^{\mu\nu}$ cannot arise in one-loop order.

The second-order contributions in $e^{\mu}$, $f^{\mu}$ and $g^{\mu\nu\lambda}$ are a little more complicated, since they involve insertions into the vertices in addition to the modification of the propagators. The graphs which illustrate these corrections are depicted below.

\vspace*{1mm}

\begin{figure}[!h]
\begin{center}
\includegraphics[angle=0,scale=1]{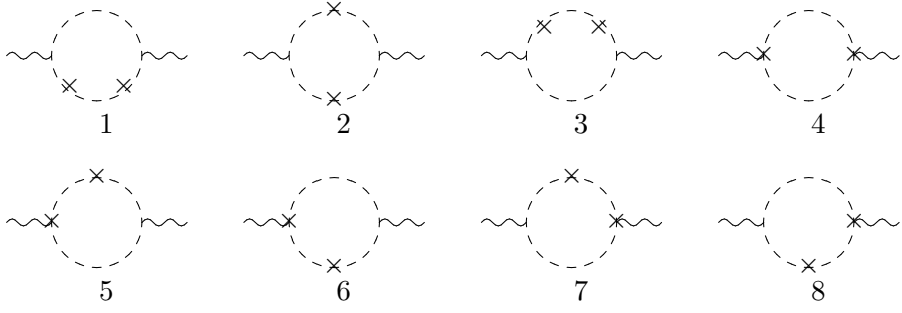}
\end{center}
\caption{The second-order contributions to the two-point function.}
\end{figure}

Let us first treat the second-order correction in $e^{\mu}$, which yields:
\begin{eqnarray}
S_{AA}^{(ae)}(p)&=&\frac{iq^2}{2}\int\frac{d^4l}{(2\pi)^4}{\rm tr}\Big(\gamma^{\alpha}\frac{1}{\ls-m}\gamma^{\beta}\frac{1}{\ls+\ps-m}
(e\cdot (l+p))\frac{1}{\ls+\ps-m}(e\cdot (l+p))\frac{1}{\ls+\ps-m}+\nonumber\\
&+&
\gamma^{\alpha}\frac{1}{\ls-m}(e\cdot l)\frac{1}{\ls-m}
\gamma^{\beta}\frac{1}{\ls+\ps-m}(e\cdot (l+p))\frac{1}{\ls+\ps-m}+\nonumber\\
&+&
\gamma^{\alpha}\frac{1}{\ls-m}(e\cdot l)\frac{1}{\ls-m}
(e\cdot l)\frac{1}{\ls-m}\gamma^{\beta}\frac{1}{\ls+\ps-m}+\nonumber\\
&+&e^{\alpha}\frac{1}{\ls-m}e^{\beta}\frac{1}{\ls+\ps-m}+\nonumber\\ &-&
e^{\alpha}\frac{1}{\ls-m}(e\cdot l)\frac{1}{\ls-m}\gamma^{\beta}\frac{1}{\ls+\ps-m}+\nonumber\\ &-&
e^{\alpha}\frac{1}{\ls-m}\gamma^{\beta}\frac{1}{\ls+\ps-m}(e\cdot (l+p))\frac{1}{\ls+\ps-m}
+\nonumber\\ &-&
\gamma^{\alpha}\frac{1}{\ls-m}(e\cdot l)\frac{1}{\ls-m}e^{\beta}\frac{1}{\ls+\ps-m}+\nonumber\\ &-&
\gamma^{\alpha}\frac{1}{\ls-m}e^{\beta}\frac{1}{\ls+\ps-m}(e\cdot (l+p))\frac{1}{\ls+\ps-m}
\Big)
A_{\alpha}(-p)A_{\beta}(p).
\label{eee}
\end{eqnarray}
A lengthy but straightforward calculation gives the gauge-invariant total result
\begin{eqnarray}
S^{(ae)}_{AA}(p)&=&\frac{q^2}{3}\left\{e^2\left(p^2\eta^{\mu\nu}-p^{\mu}p^{\nu}\right)\left(I_{log}(m^2)-\frac{1}{2\pi^2}\right)\right.
\nonumber \\&& \left. - 2 I_{log}(m^{2})\left[(e \cdot p)^{2}\eta^{\mu\nu}+p^{2}e^{\mu}e^{\nu}-(e\cdot p)(e^{\nu}p^{\mu}+e^{\mu}p^{\nu})\right]\right\}A_{\mu}(-p)A_{\nu}(p),
\end{eqnarray}
in which
\begin{equation}
I_{log}(m^2)=\int^{\Lambda}\frac{d^{4}k}{(2\pi)^{4}}\frac{1}{(k^{2}-m^{2})^2}
\end{equation}
is the basic logarithimically divergent one-loop integral defined in Implicit Regularization (see \cite{implicit} and references therein). The upper index $\Lambda$ in the integral is to indicate that the integral is regularized. Only to illustrate, if Dimensional Regularization is used, one obtains $I_{log}(m^2)=\frac{i}{16\pi^2}\Gamma(\frac{\epsilon}{2})(\frac{4\pi m^2}{\mu^2})^{\epsilon/2}$.

For the second order in $f_{\mu}$ the expression looks like
\begin{eqnarray}
S_{AA}^{(af)}(p)&=&\frac{ie^2}{2}\times\nonumber\\&\times&
\int\frac{d^4l}{(2\pi)^4}{\rm tr}\Big(\gamma^{\mu}\frac{1}{\ls-m}\gamma^{\nu}\frac{1}{\ls+\ps-m}
(f\cdot (l+p))\gamma_5\frac{1}{\ls+\ps-m}(f\cdot (l+p))\gamma_5\frac{1}{\ls+\ps-m}+\nonumber\\
&+&
\gamma^{\mu}\frac{1}{\ls-m}(f\cdot l)\gamma_5\frac{1}{\ls-m}
\gamma^{\nu}\frac{1}{\ls+\ps-m}(f\cdot (l+p))\gamma_5\frac{1}{\ls+\ps-m}+\nonumber\\
&+&
\gamma^{\mu}\frac{1}{\ls-m}(f\cdot l)\gamma_5\frac{1}{\ls-m}
(f\cdot l)\gamma_5\frac{1}{\ls-m}\gamma^{\nu}\frac{1}{\ls+\ps-m}+\nonumber\\
&+&f^{\mu}\gamma_5\frac{1}{\ls-m}f^{\nu}\gamma_5\frac{1}{\ls+\ps-m}-\nonumber\\ &-&
f^{\mu}\gamma_5\frac{1}{\ls-m}(f\cdot l)\gamma_5\frac{1}{\ls-m}\gamma^{\nu}\frac{1}{\ls+\ps-m}+\nonumber\\ &-&
f^{\mu}\gamma_5\frac{1}{\ls-m}\gamma^{\nu}\frac{1}{\ls+\ps-m}(f\cdot (l+p))\gamma_5\frac{1}{\ls+\ps-m}
+\nonumber\\ &-&
\gamma^{\mu}\frac{1}{\ls-m}(f\cdot l)\gamma_5\frac{1}{\ls-m}f^{\nu}\gamma_5\frac{1}{\ls+\ps-m}+\nonumber\\ &-&
\gamma^{\mu}\frac{1}{\ls-m}f^{\nu}\gamma_5\frac{1}{\ls+\ps-m}(f\cdot (l+p))\gamma_5\frac{1}{\ls+\ps-m}
\Big)
A_{\mu}(-p)A_{\nu}(p).
\label{eef}
\end{eqnarray}
It is obtained a purely divergent gauge-invariant result, given by
\begin{eqnarray}
S^{(af)}_{AA}(p)&=& \frac{q^2}{3}I_{log}(m^2)\left\{-f^{2}\left(p^{2}\eta^{\mu\nu}-p^{\mu}p^{\nu}\right) \right. \nonumber \\
&+&\left.2\left[(f \cdot p)^{2}\eta^{\mu\nu}+p^{2}f^{\mu}f^{\nu}-(f \cdot p)\Big(f^{\nu}p^{\mu}+f^{\mu}p^{\nu}\Big)\right]\right\}A_{\mu}(-p)A_{\nu}(p)
\end{eqnarray}

Following the same steps, we calculate the second-order correction in $g^{\mu\nu\lambda}$, given by
\begin{eqnarray}
S_{AA}^{(ag)}(p)&=&\frac{ie^2}{8}\int\frac{d^4l}{(2\pi)^4}{\rm tr}\Big(\gamma^{\mu}\frac{1}{\ls-m}\gamma^{\nu}\frac{1}{\ls+\ps-m}
\times\nonumber\\&\times&
(g^{\alpha\beta\gamma}\sigma_{\alpha\beta} (l_{\gamma}+p_{\gamma}))\frac{1}{\ls+\ps-m}
(g^{\rho\sigma\kappa}\sigma_{\rho\sigma} (l_{\kappa}+p_{\kappa}))\frac{1}{\ls+\ps-m}+\nonumber\\
&+&
\gamma^{\mu}\frac{1}{\ls-m}(g^{\alpha\beta\gamma}\sigma_{\alpha\beta} l_{\gamma})
\frac{1}{\ls-m}
\gamma^{\nu}\frac{1}{\ls+\ps-m}(g^{\rho\sigma\kappa}\sigma_{\rho\sigma} (l_{\kappa}+p_{\kappa}))
\frac{1}{\ls+\ps-m}+\nonumber\\
&+&
\gamma^{\mu}\frac{1}{\ls-m}(g^{\alpha\beta\gamma}\sigma_{\alpha\beta} l_{\gamma})\frac{1}{\ls-m}
(g^{\rho\sigma\kappa}\sigma_{\rho\sigma} l_{\kappa})
\frac{1}{\ls-m}\gamma^{\nu}\frac{1}{\ls+\ps-m}+\nonumber\\
&+&g^{\rho\sigma\mu}\sigma_{\rho\sigma}\frac{1}{\ls-m}g^{\alpha\beta\nu}\sigma_{\alpha\beta}\frac{1}{\ls+\ps-m}-\nonumber\\ &-&
g^{\rho\sigma\mu}\sigma_{\rho\sigma}\frac{1}{\ls-m}(g^{\alpha\beta\gamma}\sigma_{\alpha\beta} l_{\gamma})\frac{1}{\ls-m}\gamma^{\nu}\frac{1}{\ls+\ps-m}-\nonumber\\ &-&
g^{\rho\sigma\mu}\sigma_{\rho\sigma}\frac{1}{\ls-m}\gamma^{\nu}\frac{1}{\ls+\ps-m}(g^{\alpha\beta\kappa}\sigma_{\alpha\beta} (l_{\kappa}+p_{\kappa}))\frac{1}{\ls+\ps-m}
-\nonumber\\ &-&
\gamma^{\mu}\frac{1}{\ls-m}(g^{\alpha\beta\gamma}\sigma_{\alpha\beta} l_{\gamma})\frac{1}{\ls-m}g^{\rho\sigma\nu}\sigma_{\rho\sigma}\frac{1}{\ls+\ps-m}-\nonumber\\ &-&
\gamma^{\mu}\frac{1}{\ls-m}g^{\rho\sigma\nu}\sigma_{\rho\sigma}\frac{1}{\ls+\ps-m}(g^{\alpha\beta\kappa}\sigma_{\alpha\beta} (l_{\kappa}+p_{\kappa}))\frac{1}{\ls+\ps-m}
\Big)
A_{\mu}(-p)A_{\nu}(p).
\label{ggg}
\end{eqnarray}
Again, we have to deal with a lengthy calculation. We use the particular form of this Lorentz-breaking tensor given by $g^{\mu \nu \alpha}=\epsilon^{\mu \nu \alpha \lambda}h_\lambda$. The result is given by
\begin{eqnarray}
&&S^{(ag)}_{AA}(p)= \frac{q^2}{3}\left\{4 h^{2}I_{log}(m^2)\left(p^{2}\eta^{\mu\nu}-p^{\mu}p^{\nu}\right) + \right.\nonumber \\
&&\left.+ \left(-8 I_{log}(m^2)+\frac{i}{\pi^2}\right)
\left[(h \cdot p)^{2}g^{\mu\nu}+p^{2}h^{\mu}h^{\nu} -(h \cdot p)\Big(h^{\nu}p^{\mu}+h^{\mu}p^{\nu}\Big)\right]\right\}A_{\mu}(-p)A_{\nu}(p),
\end{eqnarray}
which, again, have a divergent part.

These second-order contributions in the parameters $e^\mu$, $f^\mu$ and $h^\mu$, as well as in the derivatives, are composed by the sum of the usual Maxwell term and the aether-like form $\kappa^{\mu\nu\lambda\rho}F_{\mu\nu}F_{\lambda\rho}$, in which the $\kappa_{\mu\nu\lambda\rho}$ tensor is written as
\begin{eqnarray}
\kappa^{\mu\nu\lambda\rho}=Q\left(u^{\mu}u^{\lambda}\eta^{\nu \rho}-u^{\mu}u^{\rho}\eta^{\lambda \nu}
-u^{\lambda}u^{\nu}\eta^{\mu \rho}+u^{\nu}u^{\rho}\eta^{\mu\lambda}\right),
\end{eqnarray}
with $Q$ being some dimensionless constant, and the role of $u_{\mu}$ being played by $e_{\mu}$, $f_{\mu}$ and $h_\mu$.

A comment is in order. Since these second-order corrections are divergent, the Lorentz-breaking terms with $e_{\mu}$, $f_{\mu}$ and $g^{\mu \nu \lambda}$ should come along with the aether term already in the classical action. It is interesting to discuss the reason why these contributions are divergent, unlike the result obtained in \cite{aether}, but in a way which is similar to the result in \cite{FerrMal}. This is related with the fact that the parameters $e^{\mu}$, $f^{\mu}$ and $g^{\mu\nu\lambda}$ are dimensionless, and, unlike in \cite{aether}, in the present case there is no unexpected cancelation of the divergence. We note that by dimensional reasons, the higher-derivative contributions from this sector will be explicitly finite.

Although it is out of the scope of this paper, we comment on the second-order corrections in $d^{\mu \nu}$, which are aether-like. The second-order correction for a particular form of $c^{\mu \nu}$ have been explicitly found in \cite{Maluf}. Indeed, the second-order contributions in $d_{\mu\nu}$, can explicitly be shown to yield exactly the same divergences as in the case of $c^{\mu\nu}$. For example, for two adjacent $d^{\mu\nu}$ vertices, one has a contribution proportional to $d^{\mu\nu}\gamma_{\mu}\gamma_5k_{\nu}(\ks-m)d^{\alpha\beta}\gamma_{\alpha}\gamma_5k_{\beta}(\ks-m)$, which, after commutation of the $\gamma_5$ matrices, yields $d^{\mu\nu}\gamma_{\mu}k_{\nu}(\ks+m)d^{\alpha\beta}\gamma_{\alpha}k_{\beta}(\ks-m)$. The UV leading contribution of this term is just the same as in the case we have $c_{\mu\nu}$ vertices instead of the $d_{\mu\nu}$ ones, that is, $c^{\mu\nu}\gamma_{\mu}k_{\nu}(\ks-m)c^{\alpha\beta}\gamma_{\alpha}k_{\beta}(\ks-m)$. Effectively, we showed that the UV leading (logarithmically divergent) contributions for both insertions are the same and can be obtained through a simple mapping of $c_{\mu\nu}\to d_{\mu\nu}$ for the second-order result in \cite{Maluf}. A similar situation occurs if the $d^{\mu\nu}$ vertices are inserted into different propagators, the only difference being that, in this case,the $\gamma_5$ matrix must be commuted not two but four times. Hence, we see that the  second-order divergent contributions in $c_{\mu\nu}$ and $d_{\mu\nu}$ are the same.

\section{Summary}

We considered the one-loop corrections to the gauge sector from the minimal Lorentz-breaking extension of the standard model. Within our calculation, we succeeded to obtain the results up to the second-order in Lorentz-breaking parameters. Up to now, this was done only within the context of the ABJ anomaly \cite{Arias} and for the $b_{\mu}$ axial vector \cite{aether}. Although experimental results put several limits on the magnitude of these parameters and, consequently, in the higher-order contributions, there are relevant aspects to be observed here. First, since the first-order corrections in some of the parameters are zero, in case of being non-null, the second-order inductions become the most important quantum contributions in these tensors. Even if the second-order corrections remain null, it is important to check if this behavior is preserved at all orders due to a deeper reason, or if it is only eventual.

Effectively, we found non-null second-order quantum corrections in $e^\mu$, $f^\mu$ and $g^{\mu \nu \alpha}$ (which are null at first-order), with some implications in the classical action. Since they furnish divergent contributions to the aether-term, we conclude that the aether-term must be introduced from the very beginning, as it is indeed done \cite{Kostel}. It is also to be noted that it can be extracted from our calculation the contributions to the renormalization constants of the minimal Lorentz-violating Standard Model up to the second-order, extending, thus, the result of \cite{Kostel}, in which the first-order contributions were found.

{\bf Acknowledgements.} This work was partially supported by Conselho
Nacional de Desenvolvimento Cient\'{\i}fico e Tecnol\'{o}gico (CNPq). The work by A. Yu. P. has been supported by the
CNPq project No. 303783/2015-0.






\end{document}